\documentclass[prl,twocolumn,showpacs,preprintnumbers,amsmath]{revtex4}

\usepackage{graphicx}
\usepackage{dcolumn}
\usepackage{bm}
\usepackage{color}

\begin{document}
\newcommand{\js}[1]
{
\textcolor{red}{{#1}}
}


\title{Surface nanobubbles: Seeing is believing}
\author{Stefan Karpitschka$^1$}
\email{stefan.karpitschka@mpikg.mpg.de}
\author{Erik Dietrich$^{2,3}$}
\email{e.dietrich@utwente.nl}
\author{James R. T. Seddon$^2$}
\email{j.r.t.seddon@utwente.nl}
\author{Harold J. W. Zandvliet$^3$}
\email{h.j.w.zandvliet@utwente.nl}
\author{Detlef Lohse$^{2}$}
\email{d.lohse@utwente.nl}
\author{Hans Riegler$^1$}
\email{Hans.Riegler@mpikg.mpg.de}
\affiliation{$^1$Max Planck Institute of Colloids and Interfaces, Department of Interfaces, Research Campus Potsdam-Golm,14476 Potsdam, Germany,\\$^2$Physics of Fluids and $^3$Physics of Interfaces and Nanomaterials,  MESA+ Institute for Nanotechnology, University of Twente, P.O. Box 217, 7500 AE Enschede, The Netherlands}

\begin{abstract}
The existence of surface nanobubbles has been previously suggested using various experimental techniques, including attenuated total reflection spectroscopy, quartz crystal microbalance, neutron reflectometry, and x-ray reflectivity, but all of these techniques provide a sole number to quantify the existence of gas over (usually) hundreds of square microns.  Thus `nanobubbles' are indistinguishable from a `uniform gassy layer' between surface and liquid.  Atomic force microscopy, on the other hand, does  show the existence of surface nanobubbles, but the highly intrusive nature of the technique means that a uniform gassy layer could break down into nanobubbles \textit{due to} the motion of the microscope's probe.  Here we demonstrate \textit{optical} visualisation of surface nanobubbles, thus validating their individual existence non-intrusively.
\end{abstract}
\pacs{}
\maketitle
Surface nanobubbles~\cite{craig2010,seddon2011} are nanoscopic gaseous objects on immersed mainly hydrophobic surfaces which can survive for days, in contrast to classical expectation.  However, up to now they have  only been individually imaged through atomic force microscopy (AFM)~\cite{ishida2000,lou2000}, whereby an immersed nanometric probe is rapidly tapped against a solid surface.  This of course leads to the obvious question: \textit{Do we see surface nanobubbles} because of \textit{the highly intrusive measurement technique, and} not \textit{because they occur naturally?}  This may seem a moot point -- the existence of surface nanobubbles can also be inferred from attenuated total reflection infra red spectroscopy (ATR-IR)~\cite{zhang2008}, quartz crystal microbalance~\cite{seo2007,zhang2008b}, neutron reflectometry~\cite{steitz2003,doshi2005}, and x-ray reflectivity~\cite{mezger2006}, so surely they do naturally occur?  The problem with all of these latter techniques is that they provide a single number to quantify the thickness of a gassy layer at the immersed near-wall region, which is a spatial average of several hundred square microns.  This is much larger than the $10^{-2}\,\mathrm{\mu m^2}$ footprint area of an individual surface nanobubble, so the results of these techniques may also be (and often is)  interpreted as a uniform gassy layer trapped between the substrate and liquid.  Bringing in a rapidly vibrating nanoscopic probe could actually lead to the break down of a uniform gassy layer into nanobubbles, as the negative pressure divergence during each retraction (often at $\sim 10^4\,\mathrm{Hz}$) causes nucleation and gas accumulation.

The necessity for new techniques that can provide evidence for nanobubble existence is clear, but every new direction taken so far has not offered individual bubble resolution and thus is open to a similar bi-interpretation as given above.  To show conclusively that surface nanobubbles naturally occur, i.e. not because of the intrusive AFM probe, we return to the good old adage of \textit{seeing is believing}.  Thus, in this Letter we use the non-intrusive technique of optical interference-enhanced microscopy~\cite{koehler2006,ohl2012} to demonstrate the existence of surface nanobubbles.

We begin by describing the experimental method.  For the substrate, we used a fluorinated silicon substrate, made hydrophobic through vapour deposition of perfluorodecyltrichlorosilane (PFDTS).  Nanobubble nucleation was achieved using the ethanol-water exchange:  Following the method of Ref.~\cite{lou2002}, the surface was initially wetted with ultra pure water in order to take a background image.  This was then replaced with ethanol as the first step in the exchange procedure.  The final step was  then to switch the liquid back to ultra pure water once more, thus allowing the proceeding exothermic chemical exchange to lead to bubble nucleation on the substrate.  In all cases, the ultra pure water was provided by a Simplicity 185 system (Millipore, France) and the ethanol was $99.9\,\mathrm{\%}$ pure (analysis grade).  In each step of the exchange the liquid volumes and flow rates were typically $1\,\mathrm{mL}$ and  $10\,\mathrm{\mu L/s}$, respectively.

\begin{figure}
\begin{center}
\includegraphics[angle=0,width=8cm]{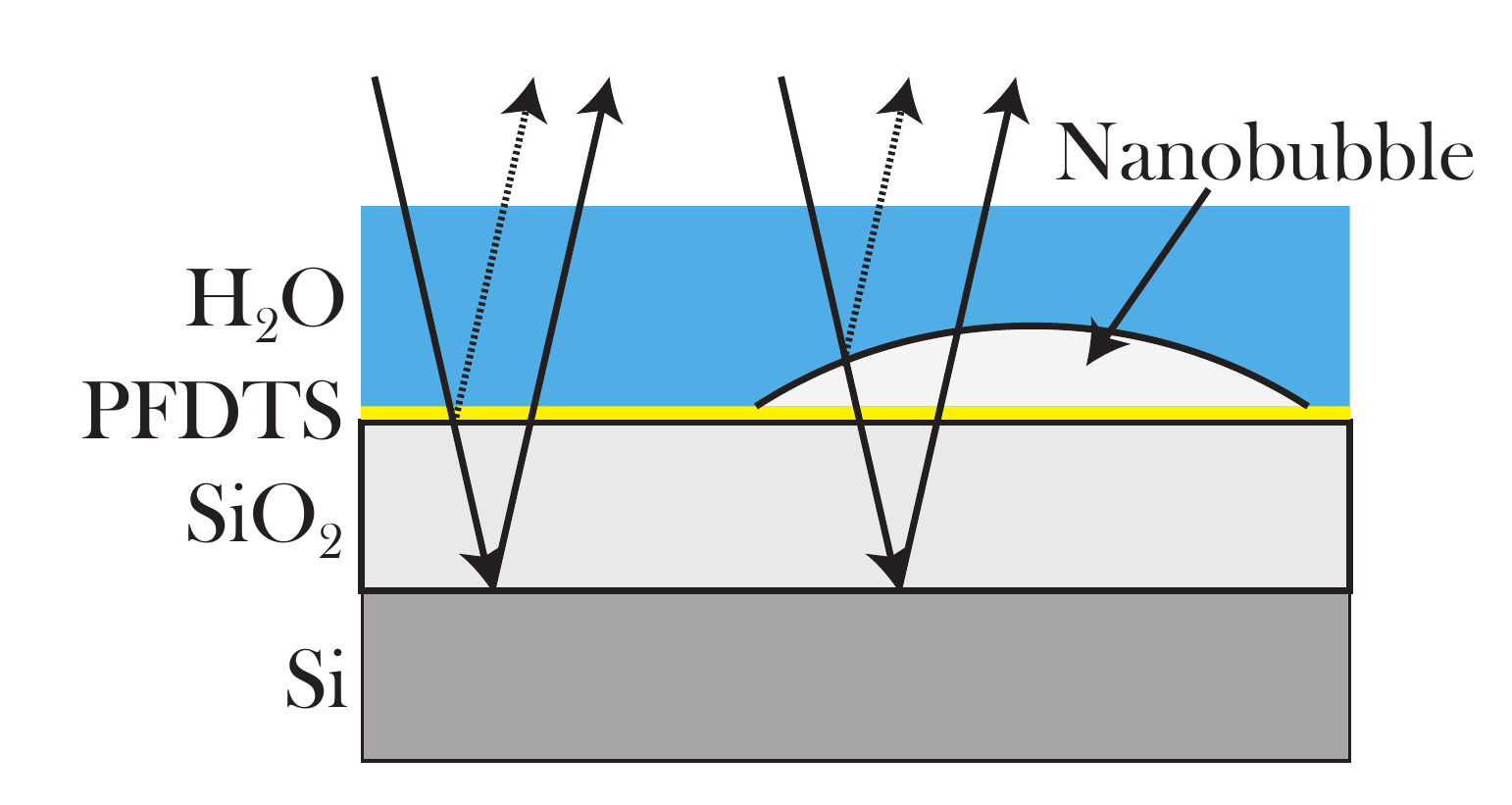}
\end{center}
\caption{\label{fig:schematic} (color online) Schematic diagram of the optical interference-enhanced microscopy technique.  The optical path length (which leads to the interference fringes) is enhanced by the double traverse through the $300\,\mathrm{nm}$ silicon oxide layer, which allows nanometrically-high near-wall layers to be visualised.}
\end{figure}

Visualisation was achieved using optical interference-enhanced microscopy~\cite{koehler2006}.  In this, the interference contrast of nanobubbles was enhanced by using silicon wafers with an artificially grown oxide layer of $300\,\mathrm{nm}$ (Silchem, Freiberg, Germany), instead of the `usual' native oxide layer of only $2\,\mathrm{nm}$. Light passes through the water, the nanobubbles, the PFDTS, and the oxide layer, before being reflected back off the silicon wafer beneath. This can then interfere with light which has been reflected directly off the top of the nanobubbles (see Fig.~\ref{fig:schematic}). By the additional oxide layer (in combination with  a properly selected wavelength), the phase difference of the interfering beams is shifted to achieve maximum contrast. The still poor contrast signal is observed by a high signal-to-noise ratio camera (Pike 100B, AVT Inc.) and further amplified by online image processing (background subtraction and oversampling, homemade software). Backround images were recorded with the substrate surface in focus, with degassed water (i.e. without nanobubbles). Images showing nanobubbles were purposefully recorded with a slightly shifted substrate ($\sim1\,\mathrm{\mu m}$), so that surface artefacts/defects are visible as dark/bright paired features.

In order to validate the objects in the optical images as surface nanobubbles, the same nucleation procedure on the same substrate was carried out within our Agilent 5100 atomic force microscope, which we operated in tapping mode using $4.5\,\mathrm{N/m}$, $10\,\mathrm{nm}$ NSC probes (MikroMasch, dry resonance of $150\,\mathrm{kHz}$).  The image size was $60\,\mathrm{\mu m} \times 60\,\mathrm{\mu m}$ to match the field of view of the optical technique.

\begin{figure*}[t!]
\begin{center}
\includegraphics[angle=0,width=15cm]{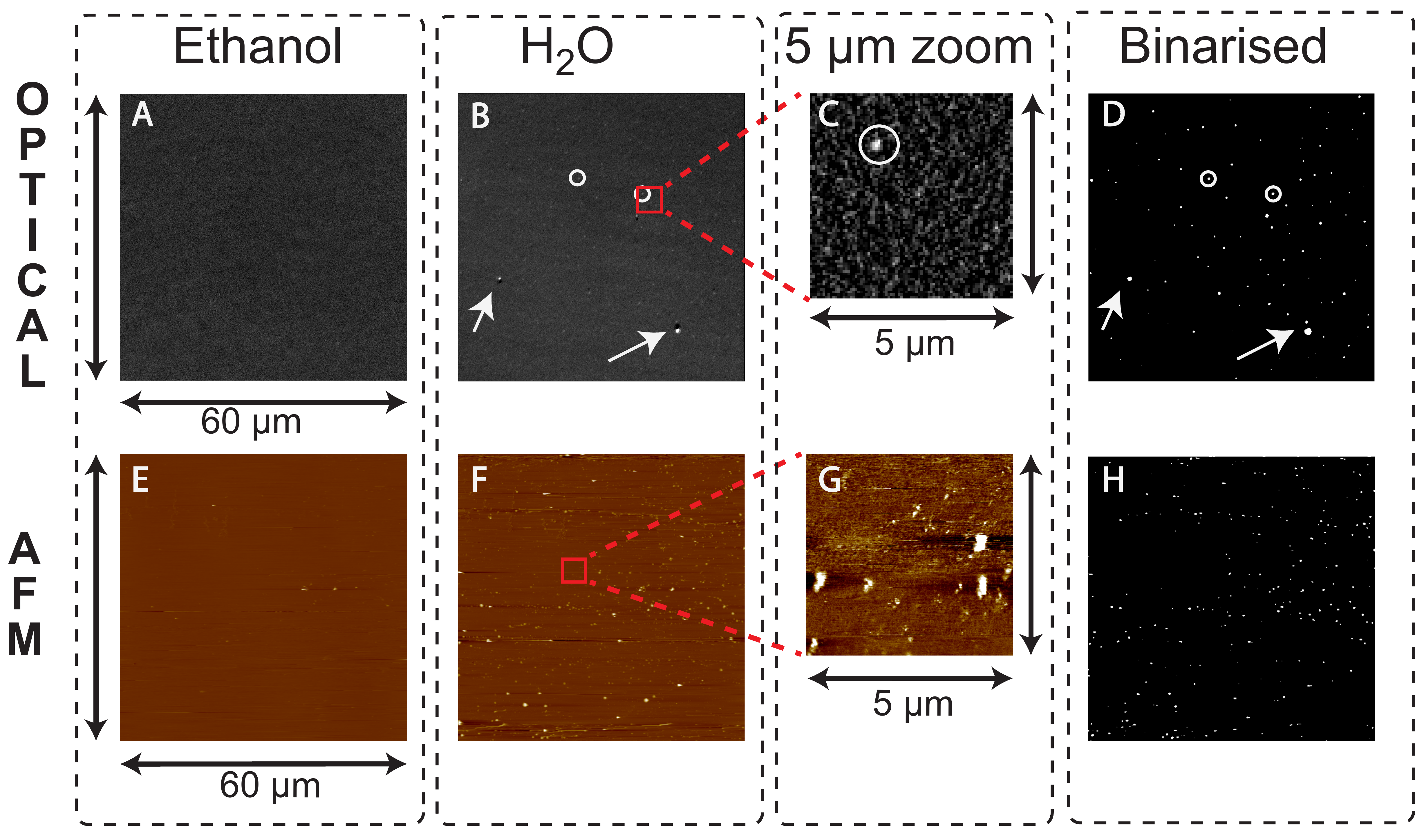}
\end{center}
\caption{\label{fig:compare}(color online) Details of the ethanol/water exchange from both the optical (A-D) and AFM (E-H) techniques.  No nanobubbles were detected in ethanol (A\&E), but several small features nucleated once the ethanol had been flushed out with water (B\&F).   C\&G are $5\,\mathrm{\mu m} \times 5\,\mathrm{\mu m}$ zoom-ins to better show the small circular features.  D\&H are binarised images of B\&F to highlight the nanobubbles on the full  $60\,\mathrm{\mu m} \times 60\,\mathrm{\mu m}$ fields of view.  The white circles are used to show the positions of some of the nanobubbles in the optical images, whilst the white arrows highlight    surface defects.}
\end{figure*}

\textit{Surface nanobubbles are optically accessible  -- } Typical images of the PFDTS surface are shown at various stages of the ethanol/water exchange in Fig.~\ref{fig:compare} from both the optical technique (Fig.~\ref{fig:compare} A--D)  and the AFM (Fig.~\ref{fig:compare} E--H).  No surface nanobubbles were detected under ethanol (Fig.~\ref{fig:compare} A\&E), which is in good agreement with the literature  (nanobubbles have only ever been found with H$_2$O as the liquid ~\cite{craig2010,seddon2011}).  As soon as water was introduced into the system and the ethanol flushed out, the surface images changed to those in Fig.~\ref{fig:compare} B\&F.  Several small circular objects had appeared, two of which are encircled in the optical image Fig.~\ref{fig:compare} B in order to highlight them from the white noise background (surface defects are also evident in Fig.~\ref{fig:compare} B, which all have `black' shadows as a result of background substraction -- we point two of these defects out with white arrows).  The surface nanobubbles are much more distinguishable in the AFM image (Fig.~\ref{fig:compare} F), which  is due to the differing levels of spatial resolution offered by the two techniques, specifically in the lower cut-off.  The optical technique is limited to objects larger than $120\,\mathrm{nm}$, whilst the current AFM image  has a lateral minimum resolution of $5\,\mathrm{\mu m}/512\,\mathrm{pixels} = 9.8\,\mathrm{nm}$.  We zoom into $5\,\mathrm{\mu m} \times 5\,\mathrm{\mu m}$ areas in Fig.~\ref{fig:compare} C\&G.  Now the optical image of the nanobubble is much clearer, again highlighted with the circle.    In order to better show the nanobubbles on the full  $60\,\mathrm{\mu m} \times 60\,\mathrm{\mu m}$ fields of view, we filter out the noise and binarise both the optical and AFM images to give Fig.~\ref{fig:compare} D\&H.  Again, it is obvious that the AFM image contains many more surface nanobubbles than the optical image (the ones below the optical resolution), but it is also clear that the AFM image contains several larger nanobubbles consistent with those seen in the optical image.

At this point we can state that the circular features detected with our optical technique appear similar to those measured with AFM using the same nucleation method on the same surface, which is also a standard nucleation method~\cite{lou2000,zhang2006c} and surface~\cite{seddon2010c} used in the nanobubble community (the results of  which are indeed interpreted as surface nanobubbles).

\begin{figure}
\begin{center}
\includegraphics[angle=0,width=8cm]{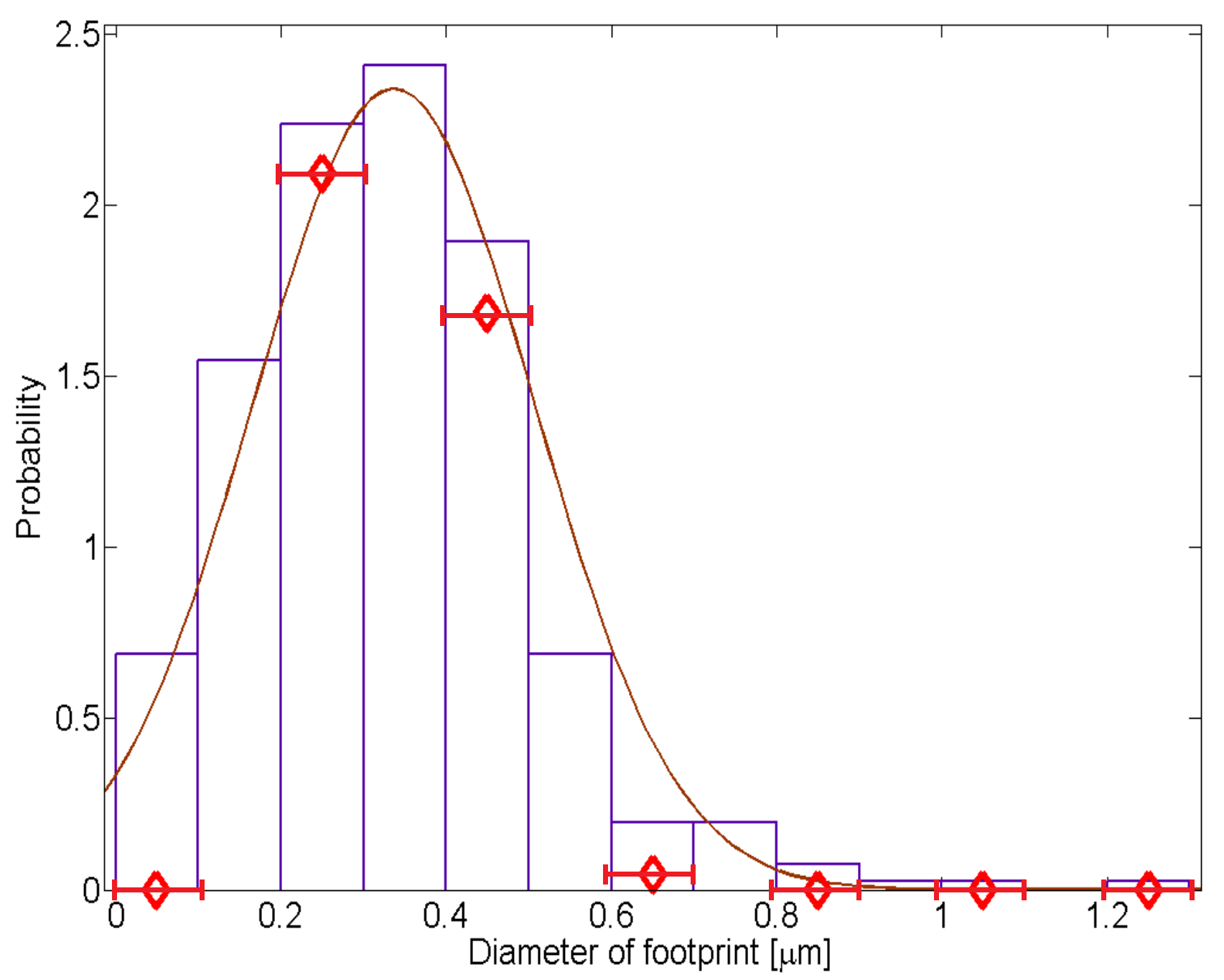}
\end{center}
\caption{\label{fig:stats1}(color online) Distributions of nanobubble footprint diameters from the AFM measurements (purple histogram and brown Gaussian fit) and optical measurements (red diamonds).   The optically-observed nanobubbles have very comparable sizes to the AFM-measured ones in the high-diameter range of applicability. The error bar indicates the $120\,\mathrm{nm}$ spatial resolution limit of the optical technique.}
\end{figure}

\textit{The optical and AFM statistics are comparable  -- } To further demonstrate that the optically-observed features are indeed surface nanobubbles, we proceed by plotting their relative size distributions in Fig.~\ref{fig:stats1} for both the optical and AFM techniques.  The AFM-measured nanobubble size distribution is Gaussian, as shown by the brown fit, with a peak corresponding to a footprint diameter of $350\,\mathrm{nm}$.  This result from the AFM measurements is in good agreement with previous statistical measures for nanobubbles~\cite{simonsen2004,borkent2009}.  The optically-measured nanobubble size distribution is also plotted in Fig.~\ref{fig:stats1}, shown as the red diamonds on top of the AFM histogram and Gaussian fit.  It is clear that the  optical distribution fits well with the AFM distribution  (the left hand side of the optical distribution is right at the limit of the spatial resolution, as shown by the horizontal error bar).  Thus, the circular objects observed with our optical technique are comparable in size and size distribution to the surface nanobubbles measured using AFM.

\begin{figure}
\begin{center}
\includegraphics[angle=0,width=8cm]{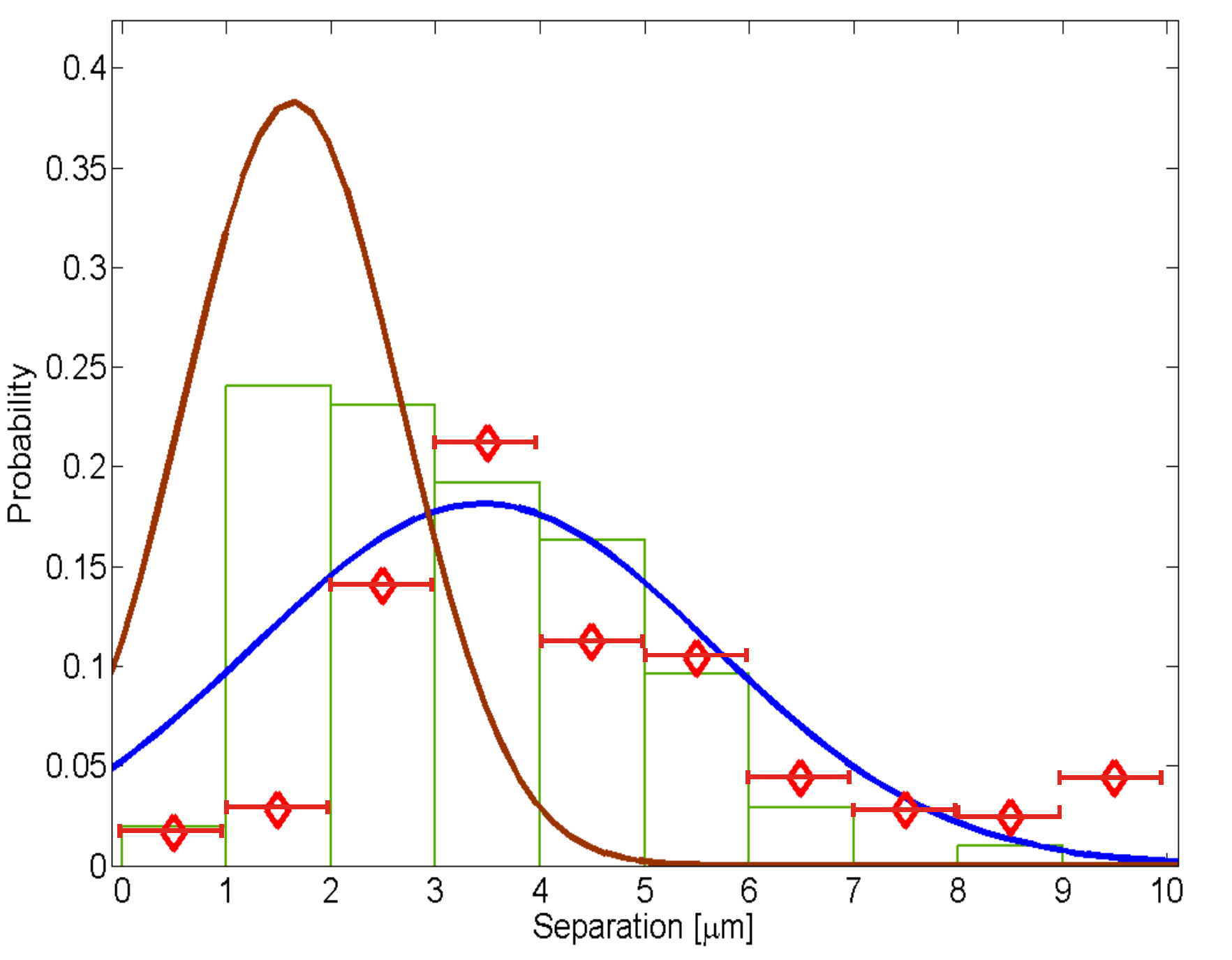}
\end{center}
\caption{\label{fig:stats2}(color online) Distributions of nearest-neighbour separations.  The brown Gaussian fit is for the entire AFM data set.  The green histogram and blue Gaussian fit is for the reduced AFM data set, considering only nanobubbles with footprint diameters $>450\,\mathrm{nm}$.  The red diamonds represent the nearest-neighbour separations for the optical measurement.  The error bar indicates $1\,\mathrm{\mu m}$, which is a good estimate for the uncertainty of the optical data.}
\end{figure}

The next statistical set that we measured was the nearest-neighbour separation, for which we plot the Gaussian fit of the AFM data as the brown curve and the optical data set as red diamonds in Fig.~\ref{fig:stats2}.  The AFM data set clearly peaks at a much lower inter-bubble separation than the optical data set, but this is due to the different minimum spatial resolutions between the two experimental techniques.  Instead, to make a more accurate comparison, we also plot the nearest-neighbour separations for a subset of the AFM data, using the exclusion rule that we only include AFM-measured nanobubbles with footprint diameters $>450\,\mathrm{nm}$, thus mimicking the minimum spatial resolution of the optical method.  This subset of the AFM data is plotted as the green histogram and blue Gaussian fit in Fig.~\ref{fig:stats2}.  It is  again clear that the optically-measured circular features are truly consistent with the AFM-measured nanobubbles.

\textit{The features are definitely surface nanobubbles -- } The surface nanobubble community has several standard tests that can be performed to confirm that the objects they image with AFMs are indeed `bubbles'.  The most obvious test is to purposefully manipulate the nanobubbles with an AFM probe, either by sweeping the nanobubbles away from a certain patch, or by coalescing neighbouring bubbles to increase their sizes.  Of course this test is impossible with the optical method -- the entire purpose of the experiment is to use a non-intrusive technique!  Instead, we performed a series of tests that could be validated against observations in the literature, which we feel further validate the optically-measured circular features as surface nanobubbles (as a direct comparison, we also carried out the tests with a solution of water and fatty acids, where the fatty acids created sub-micron droplets at the solid/liquid interface).  Details of these additional tests are listed in Table~\ref{tab:tests}.  Of these, an interesting measure was of the contact angle.  Surface nanobubbles have peculiar contact angles, in the range $\gtrsim 150\,\mathrm{^o}$~\cite{seddon2011a}, which are always much higher than  the macroscopic value (this is actually essential in order for the nanobubbles to contain Knudsen gas for their stabilising recirculating stream~\cite{seddon2011c}).  We were able to calculate the contact angles of the largest nanobubbles and  the droplets from the optical images, since these both produced several interference fringes to allow accurate measurements of their heights.  As expected, the optically-detected objects were found to have anomalously high contact angles ($\sim160\,\mathrm{^o}$) in accord with the findings of the nanobubble experiments in the literature. Surprinsgly, the contaminant droplets also had (liquid side) contact angles of $\sim160\,\mathrm{^o}$.  Thus, using contact angle alone to determine whether an object is a nanobubble or a contaminant droplet is insufficient.

\begin{table*}[width=18cm]
\caption{\label{tab:tests}Results of the further tests which we carried out to demonstrate that the optically-measured circular features were surface nanobubbles.}
\begin{tabular}{p{6cm}ccc}\hline \hline
Test & Expected  nanobubble behaviour & Observed  behaviour &   Expected droplet behaviour\\ \\ \hline
Long-time dynamics ($\sim t_0+1\,\mathrm{hr}$) & no effect~\cite{zhang2008} & no effect & Ostwald ripening\\ 
\textit{In situ} temperature increase ($\sim295+50\,\mathrm{K}$) & no effect & no effect & no effect\\
\textit{In situ} temperature decrease ($\sim345-50\,\mathrm{K}$) & no effect~\cite{berkelaar2011iti} & no effect & no effect\\ 
Temperature ramp to $373\,\mathrm{K}$ & boiling & boiling & no effect\\ 
Exchange back to ethanol & rapid dissolution ($<0.25\,\mathrm{s}$)~\cite{zhang2010b} & rapid dissolution & slow dissolution ($>1\,\mathrm{min}$)\\
Exchange to degassed water  & slow dissolution & slow dissolution & no effect\\
Contact angles\footnote{Contact angles for the largest nanobubbles were measurable, since they contained a sufficient number of interference fringes for  accurate heights to be determined} & $\gtrsim 150\,\mathrm{^o}$~\cite{seddon2011a} & $\sim160\,\mathrm{^o}$ & $\sim160\,\mathrm{^o}$\\
\hline \hline
\end{tabular}
\end{table*}

In conclusion,  we  used  interference-enhanced microscopy to optically confirm the existence of surface nanobubbles.  In order to validate the circular features indeed as being surface nanobubbles, we compared the statistics of their sizes and separations to those from atomic force microscopy measurements of a similar surface area, with very good agreement.  Furthermore, we carried out a series of validation tests on the optically-measured surface nanobubbles with comparable results to those found in the literature. 

The fact that our validation method is non-intrusive settles the on-going debate in the literature as to whether surface nanobubbles are seen in AFM measurements \textit{because of} the highly intrusive nature of the probe dynamics.  Thus, surface nanobubbles occur naturally and are not created through the breakdown of a near-wall uniform gassy layer due to an AFM probe.

Whilst the optical technique is limited to detecting only the largest surface nanobubbles, it clearly provides a new way forward to uncover nanobubble dynamics.  The method allows for both spatial and temporal resolution.

HR, SK, and ED thank Helmuth M\"ohwald for scientific advice and general support.  SK was supported by the DFG (RI1529/16-1). DL, HZ, JS, and ED acknowledge funding  from the Foundation for Fundamental Research on Matter (FOM) and the technology foundation STW, which are sponsored by the Netherlands Organization for Scientific Research (NWO).  


\begin{thebibliography}{21}
\expandafter\ifx\csname natexlab\endcsname\relax\def\natexlab#1{#1}\fi
\expandafter\ifx\csname bibnamefont\endcsname\relax
  \def\bibnamefont#1{#1}\fi
\expandafter\ifx\csname bibfnamefont\endcsname\relax
  \def\bibfnamefont#1{#1}\fi
\expandafter\ifx\csname citenamefont\endcsname\relax
  \def\citenamefont#1{#1}\fi
\expandafter\ifx\csname url\endcsname\relax
  \def\url#1{\texttt{#1}}\fi
\expandafter\ifx\csname urlprefix\endcsname\relax\def\urlprefix{URL }\fi
\providecommand{\bibinfo}[2]{#2}
\providecommand{\eprint}[2][]{\url{#2}}

\bibitem[{\citenamefont{Craig}(2011)}]{craig2010}
\bibinfo{author}{\bibfnamefont{V.~S.~J.} \bibnamefont{Craig}},
  \bibinfo{journal}{Soft Matter} \textbf{\bibinfo{volume}{7}},
  \bibinfo{pages}{40} (\bibinfo{year}{2011}).

\bibitem[{\citenamefont{Seddon and Lohse}(2011)}]{seddon2011}
\bibinfo{author}{\bibfnamefont{J.~R.~T.} \bibnamefont{Seddon}}
  \bibnamefont{and} \bibinfo{author}{\bibfnamefont{D.}~\bibnamefont{Lohse}},
  \bibinfo{journal}{J. Phys. Cond. Mat.} \textbf{\bibinfo{volume}{23}},
  \bibinfo{pages}{133001} (\bibinfo{year}{2011}).

\bibitem[{\citenamefont{Ishida et~al.}(2000)\citenamefont{Ishida, Inoue,
  Miyahara, and Higashitani}}]{ishida2000}
\bibinfo{author}{\bibfnamefont{N.}~\bibnamefont{Ishida}},
  \bibinfo{author}{\bibfnamefont{T.}~\bibnamefont{Inoue}},
  \bibinfo{author}{\bibfnamefont{M.}~\bibnamefont{Miyahara}}, \bibnamefont{and}
  \bibinfo{author}{\bibfnamefont{K.}~\bibnamefont{Higashitani}},
  \bibinfo{journal}{Langmuir} \textbf{\bibinfo{volume}{16}},
  \bibinfo{pages}{6377} (\bibinfo{year}{2000}).

\bibitem[{\citenamefont{Lou et~al.}(2000)\citenamefont{Lou, Ouyang, Zhang, Li,
  Hu, Li, and Yang}}]{lou2000}
\bibinfo{author}{\bibfnamefont{S.-T.} \bibnamefont{Lou}},
  \bibinfo{author}{\bibfnamefont{Z.-Q.} \bibnamefont{Ouyang}},
  \bibinfo{author}{\bibfnamefont{Y.}~\bibnamefont{Zhang}},
  \bibinfo{author}{\bibfnamefont{X.-J.} \bibnamefont{Li}},
  \bibinfo{author}{\bibfnamefont{J.}~\bibnamefont{Hu}},
  \bibinfo{author}{\bibfnamefont{M.-Q.} \bibnamefont{Li}}, \bibnamefont{and}
  \bibinfo{author}{\bibfnamefont{F.-J.} \bibnamefont{Yang}},
  \bibinfo{journal}{J. Vac. Sci. Technol. B} \textbf{\bibinfo{volume}{18}},
  \bibinfo{pages}{2573} (\bibinfo{year}{2000}).

\bibitem[{\citenamefont{Zhang et~al.}(2008)\citenamefont{Zhang, Quinn, and
  Ducker}}]{zhang2008}
\bibinfo{author}{\bibfnamefont{X.~H.} \bibnamefont{Zhang}},
  \bibinfo{author}{\bibfnamefont{A.}~\bibnamefont{Quinn}}, \bibnamefont{and}
  \bibinfo{author}{\bibfnamefont{W.~A.} \bibnamefont{Ducker}},
  \bibinfo{journal}{Langmuir} \textbf{\bibinfo{volume}{24}},
  \bibinfo{pages}{4756} (\bibinfo{year}{2008}).

\bibitem[{\citenamefont{Seo et~al.}(2007)\citenamefont{Seo, Yoo, and
  Jeon}}]{seo2007}
\bibinfo{author}{\bibfnamefont{H.}~\bibnamefont{Seo}},
  \bibinfo{author}{\bibfnamefont{M.}~\bibnamefont{Yoo}}, \bibnamefont{and}
  \bibinfo{author}{\bibfnamefont{S.}~\bibnamefont{Jeon}},
  \bibinfo{journal}{Langmuir} \textbf{\bibinfo{volume}{23}},
  \bibinfo{pages}{1623} (\bibinfo{year}{2007}).

\bibitem[{\citenamefont{Zhang}(2008)}]{zhang2008b}
\bibinfo{author}{\bibfnamefont{X.~H.} \bibnamefont{Zhang}},
  \bibinfo{journal}{Phys. Chem. Chem. Phys.} \textbf{\bibinfo{volume}{10}},
  \bibinfo{pages}{6842} (\bibinfo{year}{2008}).

\bibitem[{\citenamefont{Steitz et~al.}(2003)\citenamefont{Steitz, Gutberlet,
  Hauss, Kl\"osgen, Krastev, Schemmel, Simonsen, and Findenegg}}]{steitz2003}
\bibinfo{author}{\bibfnamefont{R.}~\bibnamefont{Steitz}},
  \bibinfo{author}{\bibfnamefont{T.}~\bibnamefont{Gutberlet}},
  \bibinfo{author}{\bibfnamefont{T.}~\bibnamefont{Hauss}},
  \bibinfo{author}{\bibfnamefont{B.}~\bibnamefont{Kl\"osgen}},
  \bibinfo{author}{\bibfnamefont{R.}~\bibnamefont{Krastev}},
  \bibinfo{author}{\bibfnamefont{S.}~\bibnamefont{Schemmel}},
  \bibinfo{author}{\bibfnamefont{A.~C.} \bibnamefont{Simonsen}},
  \bibnamefont{and} \bibinfo{author}{\bibfnamefont{G.~H.}
  \bibnamefont{Findenegg}}, \bibinfo{journal}{Langmuir}
  \textbf{\bibinfo{volume}{19}}, \bibinfo{pages}{2409} (\bibinfo{year}{2003}).

\bibitem[{\citenamefont{Doshi et~al.}(2005)\citenamefont{Doshi, Watkins,
  Israelachvili, and Majewski}}]{doshi2005}
\bibinfo{author}{\bibfnamefont{D.~A.} \bibnamefont{Doshi}},
  \bibinfo{author}{\bibfnamefont{E.~B.} \bibnamefont{Watkins}},
  \bibinfo{author}{\bibfnamefont{J.~N.} \bibnamefont{Israelachvili}},
  \bibnamefont{and} \bibinfo{author}{\bibfnamefont{J.}~\bibnamefont{Majewski}},
  \bibinfo{journal}{PNAS} \textbf{\bibinfo{volume}{102}}, \bibinfo{pages}{9458}
  (\bibinfo{year}{2005}).

\bibitem[{\citenamefont{Mezger et~al.}(2006)\citenamefont{Mezger, Reichert,
  Sch\"oder, Okasinski, Schr\"oder, Dosch, Palms, Ralston, and
  Honkim\"aki}}]{mezger2006}
\bibinfo{author}{\bibfnamefont{M.}~\bibnamefont{Mezger}},
  \bibinfo{author}{\bibfnamefont{H.}~\bibnamefont{Reichert}},
  \bibinfo{author}{\bibfnamefont{S.}~\bibnamefont{Sch\"oder}},
  \bibinfo{author}{\bibfnamefont{J.}~\bibnamefont{Okasinski}},
  \bibinfo{author}{\bibfnamefont{H.}~\bibnamefont{Schr\"oder}},
  \bibinfo{author}{\bibfnamefont{H.}~\bibnamefont{Dosch}},
  \bibinfo{author}{\bibfnamefont{D.}~\bibnamefont{Palms}},
  \bibinfo{author}{\bibfnamefont{J.}~\bibnamefont{Ralston}}, \bibnamefont{and}
  \bibinfo{author}{\bibfnamefont{V.}~\bibnamefont{Honkim\"aki}},
  \bibinfo{journal}{Proc. Nat. Acad. Sci.} \textbf{\bibinfo{volume}{103}},
  \bibinfo{pages}{18401} (\bibinfo{year}{2006}).

\bibitem[{\citenamefont{K\"ohler et~al.}(2006)\citenamefont{K\"ohler, Lazar,
  and Riegler}}]{koehler2006}
\bibinfo{author}{\bibfnamefont{R.}~\bibnamefont{K\"ohler}},
  \bibinfo{author}{\bibfnamefont{P.}~\bibnamefont{Lazar}}, \bibnamefont{and}
  \bibinfo{author}{\bibfnamefont{H.}~\bibnamefont{Riegler}},
  \bibinfo{journal}{Appl. Phys. Lett.} \textbf{\bibinfo{volume}{89}},
  \bibinfo{pages}{241906} (\bibinfo{year}{2006}).

\bibitem[{\citenamefont{Chan and Ohl}(2012)}]{ohl2012}
\bibinfo{author}{\bibfnamefont{C.~C.} \bibnamefont{Chan}} \bibnamefont{and}
  \bibinfo{author}{\bibfnamefont{C.-D.} \bibnamefont{Ohl}},
  \bibinfo{journal}{Phys. Rev. Lett.}  (\bibinfo{year}{2012}),
  \bibinfo{note}{submitted; in this work the authors measure growth dynamics,
  but do not validate the objects in their images as being surface nanobubbles
  rather than contaminants}.

\bibitem[{\citenamefont{Lou et~al.}(2002)\citenamefont{Lou, Gao, Xiao, Li, Li,
  Zhang, Li, Sun, Li, and Hu}}]{lou2002}
\bibinfo{author}{\bibfnamefont{S.}~\bibnamefont{Lou}},
  \bibinfo{author}{\bibfnamefont{J.}~\bibnamefont{Gao}},
  \bibinfo{author}{\bibfnamefont{X.}~\bibnamefont{Xiao}},
  \bibinfo{author}{\bibfnamefont{X.}~\bibnamefont{Li}},
  \bibinfo{author}{\bibfnamefont{G.}~\bibnamefont{Li}},
  \bibinfo{author}{\bibfnamefont{Y.}~\bibnamefont{Zhang}},
  \bibinfo{author}{\bibfnamefont{M.}~\bibnamefont{Li}},
  \bibinfo{author}{\bibfnamefont{J.}~\bibnamefont{Sun}},
  \bibinfo{author}{\bibfnamefont{X.}~\bibnamefont{Li}}, \bibnamefont{and}
  \bibinfo{author}{\bibfnamefont{J.}~\bibnamefont{Hu}},
  \bibinfo{journal}{Materials Characterization} \textbf{\bibinfo{volume}{48}},
  \bibinfo{pages}{211} (\bibinfo{year}{2002}).

\bibitem[{\citenamefont{Zhang et~al.}(2006)\citenamefont{Zhang, Maeda, and
  Craig}}]{zhang2006c}
\bibinfo{author}{\bibfnamefont{X.~H.} \bibnamefont{Zhang}},
  \bibinfo{author}{\bibfnamefont{N.}~\bibnamefont{Maeda}}, \bibnamefont{and}
  \bibinfo{author}{\bibfnamefont{V.~S.~J.} \bibnamefont{Craig}},
  \bibinfo{journal}{Langmuir} \textbf{\bibinfo{volume}{22}},
  \bibinfo{pages}{5025} (\bibinfo{year}{2006}).

\bibitem[{\citenamefont{Seddon et~al.}(2011{\natexlab{a}})\citenamefont{Seddon,
  Kooij, Poelsema, Zandvliet, and Lohse}}]{seddon2010c}
\bibinfo{author}{\bibfnamefont{J.~R.~T.} \bibnamefont{Seddon}},
  \bibinfo{author}{\bibfnamefont{E.~S.} \bibnamefont{Kooij}},
  \bibinfo{author}{\bibfnamefont{B.}~\bibnamefont{Poelsema}},
  \bibinfo{author}{\bibfnamefont{H.~J.~W.} \bibnamefont{Zandvliet}},
  \bibnamefont{and} \bibinfo{author}{\bibfnamefont{D.}~\bibnamefont{Lohse}},
  \bibinfo{journal}{Phys. Rev. Lett.} \textbf{\bibinfo{volume}{106}},
  \bibinfo{pages}{056101} (\bibinfo{year}{2011}{\natexlab{a}}).

\bibitem[{\citenamefont{Simonsen et~al.}(2004)\citenamefont{Simonsen, Hansen,
  and Kl\"osgen}}]{simonsen2004}
\bibinfo{author}{\bibfnamefont{A.~C.} \bibnamefont{Simonsen}},
  \bibinfo{author}{\bibfnamefont{P.~L.} \bibnamefont{Hansen}},
  \bibnamefont{and}
  \bibinfo{author}{\bibfnamefont{B.}~\bibnamefont{Kl\"osgen}},
  \bibinfo{journal}{J. Colloid Interface Sci.} \textbf{\bibinfo{volume}{273}},
  \bibinfo{pages}{291} (\bibinfo{year}{2004}).

\bibitem[{\citenamefont{Borkent et~al.}(2009)\citenamefont{Borkent,
  Sch\"onherr, Ca\"er, Dollet, and Lohse}}]{borkent2009}
\bibinfo{author}{\bibfnamefont{B.~M.} \bibnamefont{Borkent}},
  \bibinfo{author}{\bibfnamefont{H.}~\bibnamefont{Sch\"onherr}},
  \bibinfo{author}{\bibfnamefont{G.~L.} \bibnamefont{Ca\"er}},
  \bibinfo{author}{\bibfnamefont{B.}~\bibnamefont{Dollet}}, \bibnamefont{and}
  \bibinfo{author}{\bibfnamefont{D.}~\bibnamefont{Lohse}},
  \bibinfo{journal}{Phys. Rev. E} \textbf{\bibinfo{volume}{80}},
  \bibinfo{pages}{036315} (\bibinfo{year}{2009}).

\bibitem[{\citenamefont{{van Limbeek} and Seddon}(2011)}]{seddon2011a}
\bibinfo{author}{\bibfnamefont{M.~A.~J.} \bibnamefont{{van Limbeek}}}
  \bibnamefont{and} \bibinfo{author}{\bibfnamefont{J.~R.~T.}
  \bibnamefont{Seddon}}, \bibinfo{journal}{Langmuir}
  \textbf{\bibinfo{volume}{27}}, \bibinfo{pages}{8694} (\bibinfo{year}{2011}).

\bibitem[{\citenamefont{Seddon et~al.}(2011{\natexlab{b}})\citenamefont{Seddon,
  Zandvliet, and Lohse}}]{seddon2011c}
\bibinfo{author}{\bibfnamefont{J.~R.~T.} \bibnamefont{Seddon}},
  \bibinfo{author}{\bibfnamefont{H.~J.~W.} \bibnamefont{Zandvliet}},
  \bibnamefont{and} \bibinfo{author}{\bibfnamefont{D.}~\bibnamefont{Lohse}},
  \bibinfo{journal}{Phys. Rev. Lett.} \textbf{\bibinfo{volume}{107}},
  \bibinfo{pages}{116101} (\bibinfo{year}{2011}{\natexlab{b}}).

\bibitem[{\citenamefont{Berkelaar et~al.}(2011)\citenamefont{Berkelaar, Seddon,
  Zandvliet, and Lohse}}]{berkelaar2011iti}
\bibinfo{author}{\bibfnamefont{R.~P.} \bibnamefont{Berkelaar}},
  \bibinfo{author}{\bibfnamefont{J.~R.~T.} \bibnamefont{Seddon}},
  \bibinfo{author}{\bibfnamefont{H.~J.~W.} \bibnamefont{Zandvliet}},
  \bibnamefont{and} \bibinfo{author}{\bibfnamefont{D.}~\bibnamefont{Lohse}},
  \bibinfo{journal}{Chem. Phys. Chem.}  (\bibinfo{year}{2011}).

\bibitem[{\citenamefont{Zhang et~al.}(2010)\citenamefont{Zhang, Wu, Zhang, Li,
  and Hu}}]{zhang2010b}
\bibinfo{author}{\bibfnamefont{X.~H.} \bibnamefont{Zhang}},
  \bibinfo{author}{\bibfnamefont{Z.~H.} \bibnamefont{Wu}},
  \bibinfo{author}{\bibfnamefont{X.~D.} \bibnamefont{Zhang}},
  \bibinfo{author}{\bibfnamefont{G.}~\bibnamefont{Li}}, \bibnamefont{and}
  \bibinfo{author}{\bibfnamefont{J.}~\bibnamefont{Hu}}, \bibinfo{journal}{Int.
  J. Nanosci.} \textbf{\bibinfo{volume}{9}}, \bibinfo{pages}{383}
  (\bibinfo{year}{2010}).

\end{thebibliography}

\end{document}